\documentclass[prd,showpacs,amsmath,showkeys,floatfix,amssymb, preprintnumbers, nofootinbib, superscriptaddress]{revtex4} 
\usepackage{hyperref} 
\usepackage{amsmath,amssymb,bm}
\usepackage{epsfig}
\usepackage{graphicx,color}
\usepackage{amsfonts}
\usepackage{epstopdf}
\usepackage{cases}
\usepackage[capitalise]{cleveref}
\usepackage{color}
\usepackage{float}
\newcommand{\nn}{\nonumber}
\newcommand{\beqa}{\begin{eqnarray}}
\newcommand{\eeqa}{\end{eqnarray}}
\newcommand{\bd}[1]{ \mbox{\boldmath $#1$}  }
\begin{document}
\def\ii{\'\i}
\title{Theoretical results for hadronic masses and their widths in the framework of the SO(4) model}

\author{Tochtli Y\'epez-Mart\ii nez}
\email{tochtlicuauhtli.yepez@iems.edu.mx}
\affiliation{Instituto de Educaci\'on Media Superior de la Ciudad de
  M\'exico, Plantel General L\'azaro C\'ardenas del R\'io,
  Av. Jalalpa Norte 120. Colonia Jalalpa El Grande,
C.P. 01377, Alcald\ii a \'Alvaro Obreg\'on.  Ciudad de M\'exico,  M\'exico.}

\author{Osvaldo Civitarese}
\email{osvaldo.civitarese@fisica.unlp.edu.ar} \affiliation{
Departamento de F\'isica, Universidad Nacional de La Plata,\\
and\\
Instituto de F\'{\i}sica La Plata.
 C.C. 67 (1900), La Plata, Argentina.}

\author{Peter O. Hess} \email{hess@nucleares.unam.mx}
\affiliation{Instituto de Ciencias Nucleares, Universidad Nacional
  Aut\'onoma de M\'exico, Ciudad Universitaria,
Circuito Exterior S/N, A.P. 70-543, 04510 M´exico D.F. Mexico.\\
and\\
Frankfurt Institute for Advanced Studies, J. W. von Goethe
University, Hessen, Germany }

\author{Octavio A. Rico-Trejo}
\email{octavio.augusto@ciencias.unam.mx}
\affiliation{Instituto de Ciencias Nucleares, Universidad Nacional
  Aut\'onoma de M\'exico, Ciudad Universitaria,
Circuito Exterior S/N, A.P. 70-543, 04510 M´exico D.F. Mexico.}

\author{Ulises  I.  Ramirez-Soto}
\email{ulivinsky@ciencias.unam.mx}
\affiliation{Instituto de Ciencias Nucleares, Universidad Nacional
  Aut\'onoma de M\'exico, Ciudad Universitaria,
Circuito Exterior S/N, A.P. 70-543, 04510 M´exico D.F. Mexico.}

\date{\today}
\begin{abstract}
Abstract: The prediction of properties of the low energy portion of
the hadronic spectrum is a challenging task which, up to day, is
still tentatively given due to the non-perturbative nature of QCD at
low energies. In this paper we are exploring the validity of the
SO(4) scheme, as representative of the fundamental QCD structure of
meson-like states in the region below 2.5 GeV. We  have focussed the
attention in the calculation of the energy and width of states of
various spin, isospin and parities. The theoretical results
are compared, sistematically, to the available experimental information.
\end{abstract}


\maketitle

\section{Introduction}\label{intro}
 QCD, as the fundamental theory  of strong
interactions, is well established by now \cite{Weinberg,Lee-book}. Its high energy regime has
been explored theoretically and its predictions have been confirmed
experimentally.
In spite of the success of QCD  in predicting basic properties
of observed particles from more elementary degrees of freedom, like quarks
and gluons supplemented by the notion of confinement, the complexity
of the low energy portion of the hadronic spectrum is calling for
alternative non-perturbative approaches, like Lattice Gauge Theory (LGT)  
\cite{Dudek1,Dudek2,Dudek3,Dudek4,Dudek5,Bietenholz,Bri18,Edw20},
Dyson Schwinger Equations (DSE) \cite{Bashir,DSE-baryons,DSE-Craig1,DSE-Craig2}, and 
Coulomb gauge approaches \cite{Adam1996,Llanes2000,Adam2001,Llanes2002,Hugo2004,Hugo2011,Arturo2017}, among others. 
Based on the experience accumulated in other areas of physics, like nuclear structure
physics \cite{Ring,Bohr}, where the use of effective degrees of freedom and
symmetries has probed to be very useful at the time of avoiding
limitations in the number of variables or numerical dimensions to work with, 
we shall rely on the use of group theoretical methods to calculate the low energy portion
of the hadronic spectrum \cite{sergio1,sergio2,sergio3,sergio4}. Previous attempts along this line can be found in \cite{Hess2006,Yepez2010,SO4-1,SO4-2,SO4-3}.

Starting from the SO(4) basis \cite{SO4-4} we shall diagonalize a
Hamiltonian of the type proposed by Alvaro de Rujula \cite{Rujula}, for
states with various spin, isospin and parities. With the
resulting spectra we shall calculate mean values and energy
distributions for such states and compare our results with the
available experimental information. 
Rather than adjusting parameters for each set of states we shall 
explore the predicted density of states in order to assess 
the validity of the group theretical approach in a broad sense.

The paper is organized al follows:  The details of the calculations,
that is the structure of the SO(4) basis and the Hamiltonian, are
presented in Section \ref{so4}. The formalism used
to extract mean values and the widths for each set of meson-like
states is presented in Section \ref{energy-width}. At variance with the methods used in the analysis of the hadronic spectra (see for instance \cite{booklet2020} ) to determine the energy and width of a given state, we shall use a method which is familiar to nuclear physics analysis. It is based on Bohr and Mottelson method 
\cite{Bohr} and it consists of the choice of one state, for each spin and parity, and of the use of 
an interaction which mixes-up this state 
with the ones belonging to a background.
The analysis of the results for
meson-like states is given in Section \ref{results}.
Finally, the
conclusions are drawn in Section \ref{conclusions}.

\section{The $SO(4)$-model and mesons}
\label{so4}

In this section we shall define the elements entering the 
group structure  of the basis. The effective
quark and antiquark degrees of freedom are represented as the states
of a central potential, with orbital and spin degrees of freedom and
their interactions.

The relevant group chain for the orbital part is given by \beqa
SO(4) & \cong SU(2)_1 \otimes SU(2)_2 \supset & SO(3)
\nonumber \\
(k_1,k_2) & ~~k_1 ~~~~~~~~~~ k_2~~~~ & ~
LM
~~~.
\label{eq-1}
\eeqa
The quantities  $k_i$ ($i=1,2$) denote the quantum numbers of the
$SU_i(2)$ group, therefore the $SO(4)$ {\it irreducible
representation} (irrep) is classified by the set of
two numbers
$(k_1,k_2)$. $L$ is the orbital quantum number and $M$
is its projection. The single particle states are labelled by $k_i$
($k_1=k_2$), with $k_1$ = $0$, $\frac{1}{2}$,
$1$, $\frac{3}{2}$, like in the hidrogen atom, etc.

The present model differs significantly from the
one presented in \cite{SO4-2}, where only a two-fold
degenerate orbital state was considered, one at positive and
another one at negative energy. Here, the orbital states are treated as localized
states, as in the harmonic oscillator model
\cite{Yepez2010,Arturo2017}.

In an explicit form, the states are written as
\beqa
&&\left\arrowvert
\left.
\begin{array}{c}
k_1 ~ k_2 \\
~L_{12}~ \\
~M_{12}
\end{array}
\right\rangle
\right.\nonumber\\
&& =  \sum_{m_1m_2}(k_1m_1,k_2m_2\mid L_{12}M_{12})
\mid k_1m_1\rangle \mid k_2m_2\rangle~~.~~
\label{eq-9}
\eeqa
In the multi-quarks system, this state represents
the coupling of all quarks (antiquarks) within the $SO(4)$ scheme.
The quark and antiquark are distributed
within single-particle states, with $k_1=k_2$.
In the description of
meson states as quark-antiquark pairs, the quarks and antiquarks occupy
all possible orbitals, i.e.,
the ground state $(k_1,k_2)=(0,0)$ or the orbitals which belong to excited SO(4) representations with
$(k_1,k_2)=\left(\frac{1}{2},\frac{1}{2}\right),(1,1)$.

For baryons, one of the quarks will be
in the orbital ground state and the other two quarks will be free to
occupy any of the SO(4) configurations
$(k_1,k_2)=(0,0), \left(\frac{1}{2},\frac{1}{2}\right),(1,1)$.

Though each quark (antiquark) is in an orbital state with $k_2=k_1$,
in a multi-particle state this restriction is not present
because two $SO(4)$ irreps can be coupled to a new
SO(4)-rep $(k_3,k_4)$ with $k_3$ and $k_4$ not necessarily equal.

The angular momentum of the system $\vec{L}$, either for quark-antiquak or three quarks configurations,
is coupled to the spin $\vec{S}$ such that $\vec{J}=\vec{L}+\vec{S}$,
where $J=|L-S|,...,L+S$. Thus the
orbital configurations of the model are written as

\beqa\label{model-states}
|N_{(00)},N_{\left(\frac{1}{2}, \frac{1}{2}\right)},
N_{(1,1)},(k_1,k_2)L,S,J\rangle
~~~,
\eeqa

The quantities denoted by $N_{(k,k)}$ refer to the number of particles occupying
the orbital $SO(4)$ state $(k,k)$.
In this work we shall restrict
to quark (antiquark) flavor isospin $T_f=\frac{1}{2}$, and 
focus on specific sectors of the hadronic spectrum.

The parity of meson states, $P$, is determined by the
total orbital angular momentum $L$ , i.e.,
\beqa
P & = & (-1)^{L+1}
~~~,
\label{parity}
\eeqa
where the extra minus sign comes from the internal negative parity of the
quark-antiquark pair due to charge conjugation.

As done by De Rujula et al. \cite{Rujula}, we write for the Hamiltonian the structure:

\beqa
\small{
{\bf H}_0  = \sum_{(k_1k_1)}\omega_{(k_1k_1)}
{\bd N}_{(k_1k_1)} + a {\bd {\cal C}}_1
+ b {\bd {\cal C}}_2 + \lambda
\left( {\bd L} \cdot  {\bd S} \right)
+ \gamma {\bd S}^2
~,~}
\label{eq-h1}
\eeqa
where ${\bd N}_{(k_1k_1)}$ is the number operator of quark-anti-quark pairs in the orbital level
$(k_1k_1)$ of $SO(4)$,
${\bd {\cal C}}_k$ ($k=1,2$) are the
Casimir operators of the model,
$\left( {\bd L} \cdot  {\bd S} \right)$
is the spin-orbit interaction and
the last term is the spin-spin interaction.
The quantities  $\omega_{(k_1k_1)}$ are the orbital energies of a quark-antiquark pair in the 
levels denoted by $(k_1,k_1)$.
It is well known that in order to describe confined quarks a linear potential
should be added to the Coulomb term. It is then expected that the orbital energies of 
confined quarks would resemble those of a localized central potential, namely:
\beqa
\omega^H_{(k_1,k_1)} & = & \omega_{00}^H
+\left(1-\frac{1}{(2 k_1 +1)^2}\right)\omega^H_{00}
~~~.
\label{energy-hydrogen}
\eeqa
as well as those belonging to a linear potential
\beqa
\omega^L_{(k_1,k_1)} & = & \omega^L_{(0,0)} (2 k_1 +1)
~~~.
\label{linear-energies}
\eeqa

In the following we shall denote with  {\it Model H} and
{\it Model L} both types of orbital energies.

The schematic nature of the model is enforced by the conservation of the flavor  symmetry, i.e.,
the values $\omega_{(k_1k_1)}$ do not distinguish between
up and down quarks, and by keeping the degeneracy of the hypercharge ($Y$) and the isospin ($T$) sectors.
e.g: Gel'man-Okubo terms are not included in the Hamiltonian.

For the multi-quark (quark-antiquark or three quarks) states,
the  $SO(4)$ representations are  labelled by $(k_1,k_2)$.
The eigenvalues of the Casimir operators are
\cite{Greiner, SO4-1,SO4-2,SO4-3,SO4-4}

\beqa
{\bd {\cal C}}_1 & \rightarrow & k_1 (k_1+1)+k_2(k_2+1)
\nonumber \\
{\bd {\cal C}}_2 & \rightarrow & k_1 (k_1+1)-k_2(k_2+1)
~~~.
\label{eq-h2}
\eeqa
In particular, the eigenvalue of the operator ${\bd {\cal C}}_2$ vanishes when
$k_1=k_2$, but for any other SO(4) representation the Casimir operators
generate a richer structure in the spectrum of the Hamiltonian of Eq. (\ref{eq-h1}).

In the section \ref{energy-width}
we shall focus on specific model configurations,
Eq. (\ref{model-states}): with quantum numbers $J^P$ (angular momentum and parity) of
meson states, i.e., $J^P=0^\pm,1^\pm,2^\pm$.

The Hamiltonian of Eq.(\ref{eq-h1}) has a set of five parameters,
$\{\omega_{00}, a,b,\lambda, \gamma\}$. In order to determine the
value of each of these parameters we have chosen five representative
masses, belonging to the low, medium and high energy regions of the
meson spectrum, for each of the spin and parity values. The set of representative states is shown in  
Table \ref{tab-m1}, where we are given the quantum numbers and masses of these states. The
values obtained for the parameters for each of the models of Eq.
(\ref{energy-hydrogen}) and (\ref{linear-energies}) are given in Table\ref{tab-m2}. The quark
(antiquark) isospin-symmetry is considered as exact, and for
simplicity we restrict to  $u$ and $d$ quarks, with no flavor
interactions, and to quark-antiquark states with zero color. Because of the
two-flavor representation each meson state of the model is {\it
4-fold degenerate}. The $s$-quark degree of freedom is also
excluded. It is noted that the values of the scales $\omega^H_{00}$ and $\omega^L_{00}$
given in  Table \ref{tab-m2} are almost independent of the quantum numbers $J^P$
(spin and parity) of the meson states.

\begin{center}
\begin{table}[h]
\centering
\begin{tabular}{|c|c||c|c||c|c|}
\hline
mesons &Energy\\
$J^P$ & [MeV] \\
\hline
$0^-$ & 500  \\
$0^+$ & 800  \\
$1^-$ & 900  \\
$1^+$ & 1270  \\
$2^-$ & 1900  \\
\hline
 \end{tabular}
\caption{Energy of the meson states used to extract the parameters of the model Hamiltonian. }
\vspace{0.2cm}
\label{tab-m1}
\end{table}
\end{center}

\begin{center}
\begin{table}[h]
\centering
\begin{tabular}{|c|c||c|c||c|c|}
\hline
Model H & Value & Model L & Value \\
Parameters & [MeV]& Parameters & [MeV] \\
\hline
$\omega^H_{00}$ & 250.0 & $\omega^L_{00}$ & 250.0  \\
$a^H$ & 388.3     & $a^L$ & 346.6  \\
$b^H$ &  -251.6   & $b^L$ &  -303.3 \\
$\lambda^H$ &374.1  & $\lambda^L$ &343.3  \\
$\gamma^H$ & 200.0 & $\gamma^L$ & 200.0  \\
\hline
 \end{tabular}
\caption{Parameters extracted from the fit to the energies of the meson states shown in Table \ref{tab-m1}. The columns labelled Model H and Model L are the values extracted by using the hydrogen-atom-like and the linear-potential ansatz for $\omega(k_1,k_1)$, respectively. These values have been obtained in Ref.\cite{SO4-4}}
\vspace{0.2cm}
\label{tab-m2}
\end{table}
\end{center}

\section{Mean value of the energy and the width of mesonic states.}
\label{energy-width}
The schematic Hamiltonian discussed in the previous section provides the reference states for each meson-like states of
 a given angular momentum and parity. Then, one should add to it an interaction in order to redistribute the strength of the reference states by mixing them with states belonging to a given subspace. We may define the matrix elements of the interaction by means of the coupling scheme

\beqa\label{v-interaction}
&&V_{a \alpha}\nonumber\\
&&=V_0  \langle k_1 M_{k_1}, k_2 M_{k_2} | L_a M_{L_a} \rangle
\langle L_a M_{L_a}, S_a M_{S_a} | J M_J \rangle
\langle k_3 M_{k_3}, k_4 M_{k_4} | L_\alpha M_{L_\alpha} \rangle
\langle L_\alpha M_{L_\alpha}, S_\alpha M_{S_\alpha} | J M_J \rangle
\eeqa
where the sub-index $a$ labels a reference state and the sub-index $\alpha$ denotes any state of the subspace which can couple with the state $a$. One usually talks of $\alpha$ as a state belonging to the background. Thus, the interaction term (V ), describes the interactions not included in the SO(4) scheme.

The Hamiltonian is written
\beqa\label{H}
\bf{H}=\bf {H}_0+\bf{V}
\eeqa
where $\bf{H}_0$ is the $SO(4)$ Hamiltonian of Eq. (\ref{eq-h1}).
To calculate the width of a state defined by the set of numbers, as given in Eq. (\ref{model-states}).

We assume that the basis can then be written as a set of reference states ${|a\rangle}$ and a
background ${|\alpha\rangle}$ such that
\beqa\label{elements}
{\bf{H}_0} | a \rangle &=& E_{a} | a \rangle\nonumber\\
{\bf{H}_0} | \alpha \rangle &=& E_{\alpha} | \alpha \rangle\nonumber\\
\langle a | {\bf{V}} | a \rangle &=& 0\nonumber\\
\langle \alpha_j | {\bf{V}} | \alpha_{j'} \rangle &=& V_{ \alpha_{j}, \alpha_{{j'} }}=0
~~~~\forall ~j,j' \nonumber\\
\langle a | {\bf{V}} | \alpha_{j} \rangle &=& V_{a,\alpha_j}=V_{\alpha_j,a}
=\mbox{real} ~.
\eeqa
leading to the Hamiltonian matrix
\beqa\label{matrix}
\left( \begin{array}{cccccc}
E_{a} & V_{a,\alpha_1}     & V_{a,\alpha_2}& V_{a,\alpha_3} &\cdots&V_{a,\alpha_N} \\
V_{a,\alpha_1} & E_{\alpha_1} & 0                        &0                        & \cdots&0          \\
V_{a,\alpha_2} & 0                         &E_{\alpha_2} & 0                       &\cdots&0\\
 . &. &.&.&.\\
V_{a,\alpha_N} & 0                        &0                         &0                          &\cdots&E_{\alpha_N} \\
\end{array} \right) \, ,
\eeqa

Any eigenstate of the Hamitonian of Eq. (\ref{H}) can be written as
\beqa\label{eigenstate}
| E \rangle = c_{a}(E) | a\rangle
+\sum_{j} c_{\alpha_j}(E)  | \alpha_j\rangle ~.
\eeqa

The above equations and the normalization condition $\langle E \mid E\rangle=1$ lead to the amplitudes
\beqa\label{amps}
c_{\alpha_j}(E)  &=&- c_{a}(E) \frac{ V_{a,\alpha_j} } { (E_{\alpha_j}-E)} \nn\\
\left( c_{a}(E)  \right)^2
&=&
\left( 1+ \sum_{j} \frac{ \left( V_{ a,\alpha_j } \right)^2}{ ( E_{\alpha_j }-E )^2 }\right)^{-1}~.
\eeqa

Then, the mean value for the energy and the width of the
$| a\rangle$ state when $E \approx E_a$ are
\beqa\label{mean}
\bar E &=& E_a \left(c_a(E)\right)^2 +\sum_j  E_{\alpha_j}\left(c_{\alpha_j}(E)\right)^2\nn\\
\Gamma&=&2\sigma\nonumber\\
&=&2 \left( (E_a - \bar E)^2 \left(c_a(E)\right)^2
+\sum_j  (E_{\alpha_j}-\bar E)^2\left(c_{\alpha_j}(E)\right)^2
\right)^\frac{1}{2}\nonumber\\
\eeqa

\section{Results: Mean value and width of the states.}
\label{results}

In this section we are presenting the results of the systematics, about energies and widths of the mesonic states.
For each set of mesons with 
quantum numbers $J^{P}$ we shall present and discussed the results for the calculated spectrum and 
widths and make a comparison with the available data. All data are taken from the Particle Data Group booklet \cite{booklet2020}.

It is worth to mention that the general strategy is to fix the strength of the interaction of Eq. (\ref{v-interaction}), which is the same for all states. Its value is fixed at $V_0=100 $ MeV. The results are then mostly predictions. We expect to retrieve gross properties of the spectrum, because we do not break the flavor symmetry nor we add the Gellman-Okubo terms to the Hamiltonian. The comparison will be done to the gross properties of the spectrum, for each spin and parity. At the end, the total number of states will be cast into a single figure, comparing the theoretical results with the experimental ones.

\subsection{Mesons $0^+$}

\begin{figure}[H]
\centering
\includegraphics[width=0.7\textwidth]{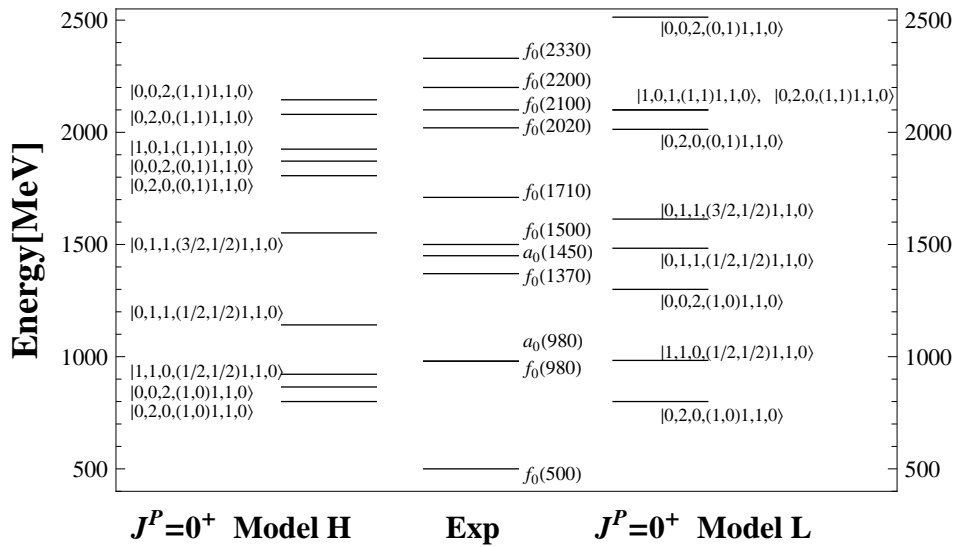}
\caption{Experimental and theoretical values for $J^P=0^+$ mesons. The set of numbers defined in Eq.(\ref{model-states}) is shown for each of the theoretical model Hamiltonians, denoted by Model H and Model L, respectively. The column Exp shows the experimental values.}
\label{HLMesones0P}
\end{figure}

Let us first discuss the case of mesons with 
 quantum numbers $J^P=0^+$. The identification of low-energy mesons with these quantum numbers 
is very controversial, because of the large widths of some of these states.
Table \ref{tab-m-0+} is a list of $J^P=0^+$ mesons, their masses and widths. The experimental values and the theoretical results for states belonging to this subspace of meson-like states are shown in Figure \ref{HLMesones0P}.

\begin{table}[H]
\centering
\caption{Experimental values for mesons with $J^P$ = $0^{+}$, $I$ denotes the isospin of the states. The first column shows the experimental value of the energy and the third one the width of the states. In the last three colums are listed the SO(4) numbers of the reference states, the mean value of the calculated energies and the theoretical value of the widths.The numbers shown in the fourth columb are those of the reference states corresponding to the L-model.}
\label{tab-m-0+}
\begin{tabular}{|c|c|c|c|c|c|}
\hline
~E[MeV]~&$I^{G}(J^{PC})$&width ([MeV])
&$|N_{(00)},N_{\left(\frac{1}{2}, \frac{1}{2}\right)},
N_{(1,1)},(k_1,k_2)L,S,J\rangle$ & $\bar{E}$[MeV] & $\Gamma$[MeV] \\ \hline

$f_{0}(500)$&$0^{+}(0^{++})$&$400-700$       
&$| 0, 2, 0, (1, 0), 1, 1, 0 \rangle $      & 808.55       &  151.60  \\

$f_{0}(980)$&$0^{+}(0^{++})$&$10-100$
&$| 1, 1, 0, (\frac{1}{2}, \frac{1}{2}), 1, 1, 0 \rangle $     & 984.37     &  113.98  \\

$a_{0}(980)$&$1^{-}(0^{++})$&$50-100$
&&&\\

$f_{0}(1370)$&$0^{+}(0^{++})$&$200-500$
&$| 0, 0, 2, ( 1, 0), 1, 1, 0 \rangle $ & 1304.21       &  152.86 \\

$a_{0}(1450)$&$1^{-}(0^{++})$&$265$ 
&&&\\
$f_{0}(1500)$&$0^{+}(0^{++})$&$112$
&$| 0, 1, 1, (\frac{1}{2}, \frac{1}{2}), 1, 1, 0 \rangle $ & 1483.53      &  113.27  \\

$f_{0}(1710)$&$0^{+}(0^{++})$&average: $147$, used value: $123$
&&&\\
$a_{0}(1950)$&$1^{-}(0^{++})$&$271$
&&&\\
$f_{0}(2020)$&$0^{+}(0^{++})$&$442$
&$| 0, 2,0, (0,1), 1, 1, 0 \rangle $ & 2019.68      &  144.53  \\

$f_{0}(2100)$&$0^{+}(0^{++})$&$284$
&$| 0, 2,0, (1,1), 1, 1, 0 \rangle $ & 2096.38      &  68.48  \\

&&&$| 1, 0,1, (1,1), 1, 1, 0 \rangle $ & 2097.14      &  80.27  \\

$f_{0}(2200)$&$0^{+}(0^{++})$&$207$
&&&\\
$f_{0}(2330)$&$0^{+}(0^{++})$&$149-223$
&&&\\ \hline
\end{tabular}
\end{table}

In spite of the large measured widths, for some of the states, the present model seems to be able to
reproduce several characteristics of this subspace, like the density of states and the sequence of energies up to 2.5 GeV. The energy separation between  states is larger for the Model L but the pile-up of states is better reproduced by the results obtained with Model H.

\begin{figure}[H]
\centering
\includegraphics[width=0.75\textwidth]{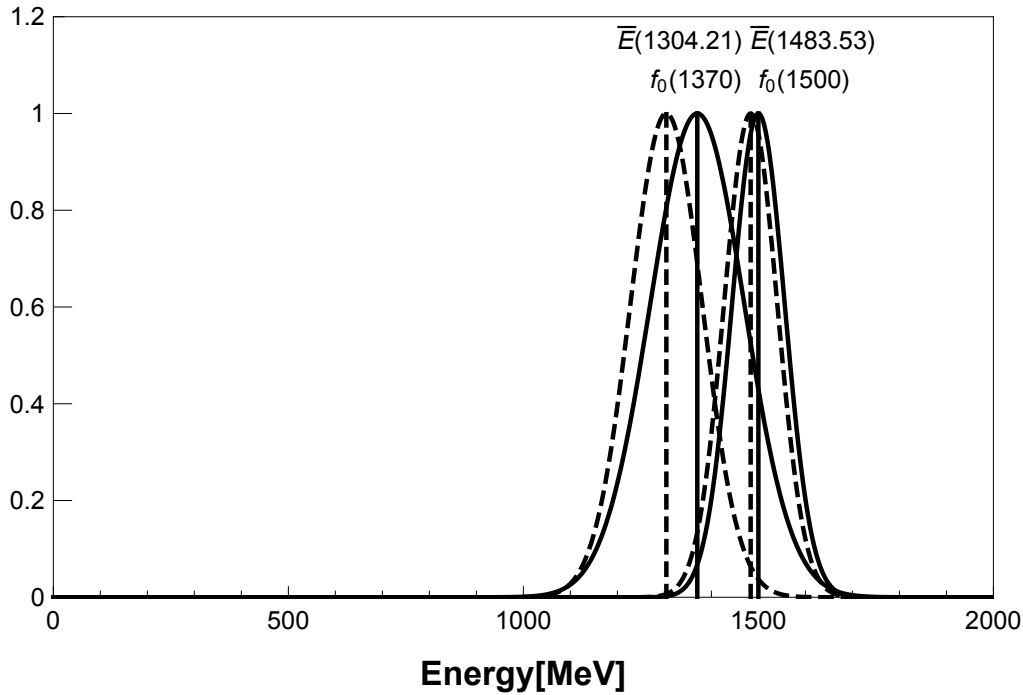}
\caption{Gaussian plots of the experimental energies and widths (solid lines)  of the states $f_0(1370)$ and $f_0(1500)$ and their theoretical values (dashed-lines).}
\label{GPf034v2}
\end{figure}

An example of the correspondence between data and theoretical results, for these mesons, is given in Figure \ref{GPf034v2}, where the experimental and theoretical values for $f_0(1370)$ and $f_0(1500)$ states are shown. The calculated values of the mean-energy agree quite well with the experimental values.

Our theoretical results point in the direction of two states with widths of the order of $100-200 MeV$.
As it is seen in Table \ref{tab-m-0+}, in the range of energy of these experimental and theoretical states, the $a_0(1450)$ state is  reported, however a gaussian plot for this state and its corresponding width will wrap both the experimental and theoretical states. This is also in relatively good agreement with the theoretical results since the multiplets $\bar{E}(1304.21)$ and $\bar{E}(1483.53)$ also includes states with isospin one. 

A similar analysis was performed for the other meson-like states of different spin and parity. The results of the complete set of calculations are shown in Figure \ref{Widths}

\subsection{Mesons $0^-$}
\label{mesons}

\begin{figure}[H]
\centering
\includegraphics[width=0.7\textwidth]{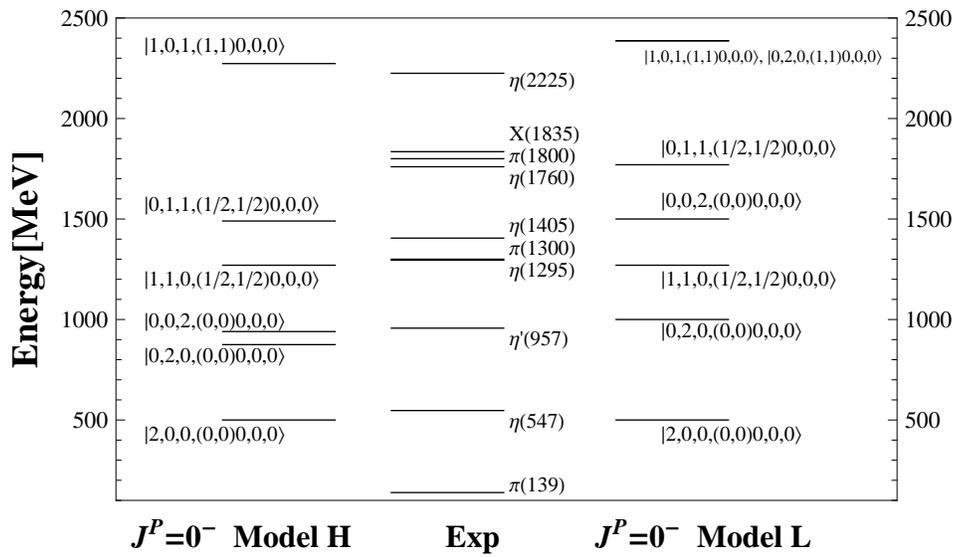}
\caption{Theoretical (Model H and Model L) and experimental (Exp) values for pseudoscalar mesons with $J^P=0^-$.}
\label{HLMesones0M}
\end{figure}

For the pseudoscalar meson spectrum $J^P=0^-$,  the data
consists of ten states with masses in the range
$139-2225$ MeV. The theoretical assumptions, concerning the adoption of the set of numbers associated to reference states, for this case, are shown in the last two columns of Table \ref{tab-m-0-}.
The pseudoscalar meson spectrum is shown in Fig. \ref{HLMesones0M}. This
subspace of the whole meson spectrum is dominated by
$\eta$-states. However, it seems that almost at every energy where an
$\eta$-state is reported, there is also a
$\pi$-state. It is worth to mention that we have taken as reference state the two states  which exhibit the larger width, as indicated in the forth column of Table \ref{tab-m-0-}. The other states have smalller spreading widths,

It is seen that certain
characteristics of the spectrum are well described in the context of the SO(4) representation. It is noted that the 
energy spacing in {\it Model-L} is larger than the energy spacing obtained with Model H, but still the results of both models  
are somehow similar.

\begin{center}
\begin{table}[h]
\centering
\begin{tabular}{|c|c|c|c|c|c|}
\hline
~E[MeV]~&$I^{G}(J^{PC})$&width ([MeV])
&$|N_{(00)},N_{\left(\frac{1}{2}, \frac{1}{2}\right)},
N_{(1,1)},(k_1,k_2)L,S,J\rangle$ & $\bar{E}$[MeV]   &  $\Gamma$[MeV] \\ \hline
$\pi^{0}(134)$&$1^{-}(0^{-+})$&$-$
&&&\\
$\pi^{\pm}(139)$&$1^{-}(0^{--})$&$-$
&&&\\
$\eta(547)$&$0^{+}(0^{-+})$&$1.31$keV
&&&\\
$\eta '(958)$&$0^{+}(0^{-+})$&$0.188$
&&&\\
$\eta(1295)$&$0^{+}(0^{-+})$&$55$
&&&\\
$\pi(1300)$&$1^{-}(0^{-+})$&$200-600$
&$|1, 1, 0, (\frac{1}{2}, \frac{1}{2}), 0, 0, 0\rangle $ &  1272.75    &  265.17  \\
$\eta(1405)$&$0^{+}(0^{-+})$&$50.1$
&&&\\
$\eta(1475)$&$0^{+}(0^{-+})$&$90$
&&&\\
$\eta(1760)$&$0^{+}(0^{-+})$&$240$
&$|0, 1, 1, (\frac{1}{2}, \frac{1}{2}), 0, 0, 0\rangle $ &  1748.03    &  264.00 \\
$\pi(1800)$&$1^{-}(0^{-+})$&$215$
&&&\\
$X(1835)$&$?^{?}(0^{-+})$&$242$
&&&\\
$\eta(2225)$&$0^{+}(0^{-+})$&$185$
&&&\\ \hline
 \end{tabular}
\caption{Experimental energies for mesons
with $J^P$ = $0^{-}$ and isospin $I$ (first and third columns) and their theoretical values (last two colums).}
\label{tab-m-0-}
\end{table}
\end{center}

\begin{figure}[H]
\centering
\includegraphics[width=0.7\textwidth]{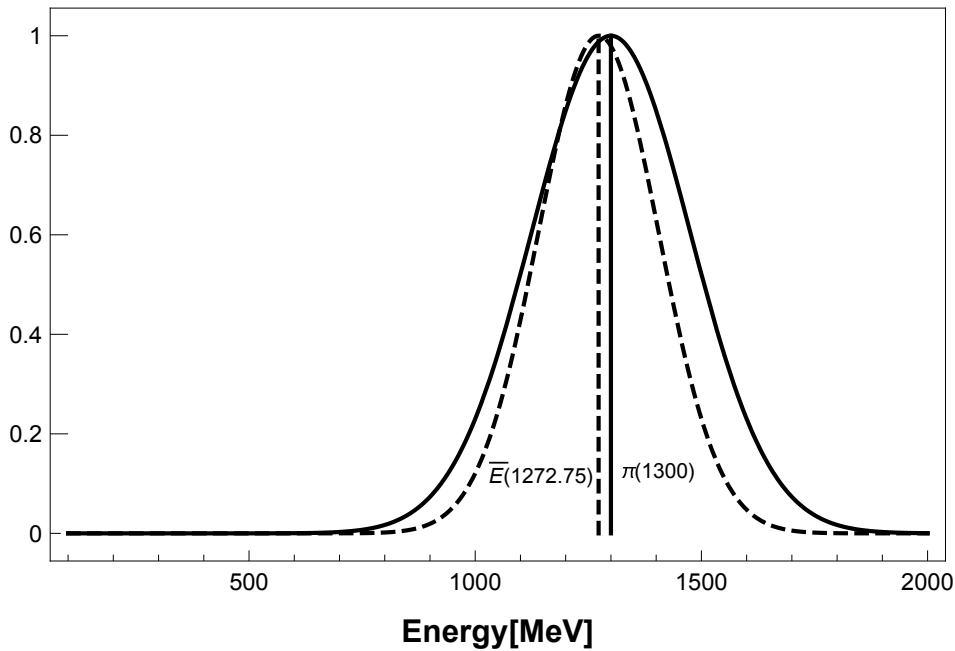}
\caption{Gaussian distribution plots of $J^P=0^-$ meson states 
$(\eta(1295),\pi(1300))$. Their energies and  widths (solid lines) are compared to the calculated value $\bar{E}(1272.75)$ and its theoretical width (dashed line).}
\label{GPpi-eta-T_1300}
\end{figure}

In Figure \ref{GPpi-eta-T_1300},  the theoretical multiplet $\bar{E}(1272.75)$ is compared to the experimental $(\eta(1295),\pi(1300))$ states, which we have associated as an isospin multiplet according to their energies. It is observed that the width of the experimental state $\pi(1300)$ is slightly larger than the theoretical one. This could be due to the fact that from the experimental side there are more decay channels reported, each with a different value of the width, while on the theoretical side there are a few states interacting with the  multiplet taken as the reference state.Therefore, the theoretical multiplet associated to this energy domain has the quantum numbers of the  $(\eta(1295),\pi(1300))$-states, an energy very close too  and a width which includes features of both $(\eta(1295),\pi(1300))$-states.

We have also compared the theoretical multiplet at 1770 MeV with the experimental $(\eta(1760),\pi(1800))$ states. It is observed (see Table \ref{tab-m-0-}) that the experimental widths are very similar to the calculated ones. The mean energy $\bar{E}(1748.03)$  and its width $\Gamma=264 MeV$ both agree quite nicely with the experimental ones, considering the simplicity of the model.

\subsection{Mesons $1^-$}

\begin{figure}[H]
\centering
\includegraphics[width=0.7\textwidth]{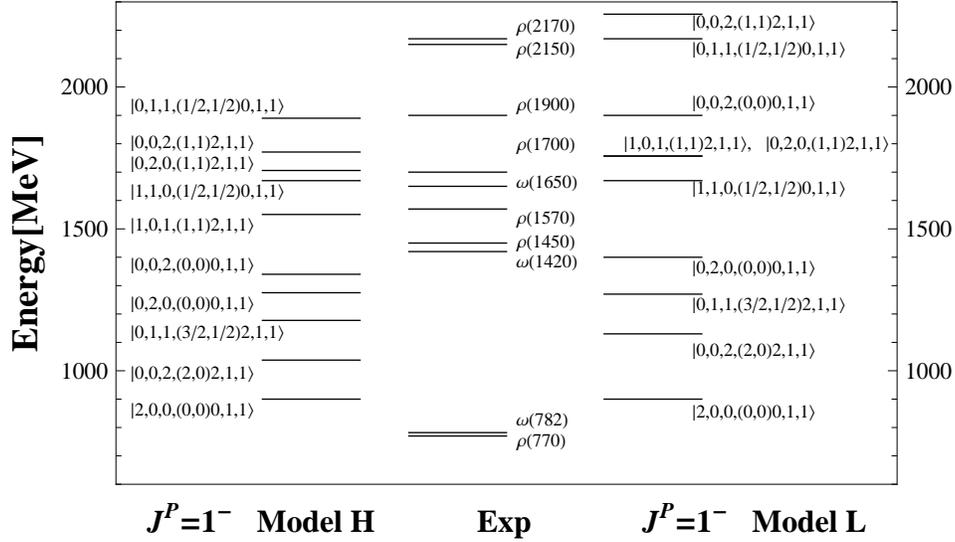}
\caption{Experimental and theoretical values of the energy for $\rho$ and $\omega$ mesons. The values are shown as explained in the captions to Figure 1.}
\label{HLMesones1M}
\end{figure}

The subspace of vector mesons contains the
$\rho$ and $\omega $
states.
The ground state, in the quark model, restricts
to isospin 0 and 1, respectively, thus only the $\rho$ and $\omega$ mesons satisfy this exact symmetry.
Therefore, we accommodate them into multiplets according to their energies. The comparison between experimental and theoretical values is shown in Table (\ref{tab-m-1-}), where the reference states are listed, and in Figure  \ref{HLMesones1M}.

\begin{table}[H]
\centering
\caption{Experimental spectrum for mesons
with $J^P$ = $1^{-}$ and isospin $I$ and their theoretical values. The values are depicted as explained in the captions to Table III.}
\label{tab-m-1-}
\begin{tabular}{|c|c|c|c|c|c|}
\hline
E [MeV]&$I^{G}(J^{PC})$&width ([MeV])
&$|N_{0},N_{\frac{1}{2}},N_{1},(k_{1},k_{2}),L,S,J\rangle$ & $\bar{E}$[MeV]  & $\Gamma$[MeV] \\  \hline
$\rho(770)$&$1^{+}(1^{--})$&$147.8$
&$|2, 0, 0, (0, 0), 0, 1, 1\rangle$ &  946.04      &  324.90  \\
$\omega(782)$&$0^{-}(1^{--})$&$8.49$
&&&\\ 
&&&&&\\

$\omega(1420)$&$0^{-}(1^{--})$&$290$
&&&\\
$\rho(1450)$&$1{+}(1^{--})$&$400$
&$|0, 2, 0, (0, 0), 0, 1, 1\rangle$ &  1399.75      &  336.65  \\

$\rho(1570)$&$1^{+}(1^{--})$&$144$
&$|1, 1, 0, (\frac{1}{2}, \frac{1}{2}), 0, 1, 1\rangle$ &  1671.67      &  259.31  \\

$\omega(1650)$&$0^{-}(1^{--})$&$315$
&$|1, 0, 1, (1, 1), 2, 1, 1\rangle$ &  1755.59      &  71.28  \\
$\rho(1700)$&$1{+}(1^{--})$&$250$
&$|0, 2, 0, (1, 1), 2, 1, 1 \rangle$&  1756.44      &  79.97  \\
&&&&&\\

$\rho(1900)$&$1^{+}(1^{--})$&The reported values are in 
&&&\\
&& the range of $10-160$
&&&\\
$\rho(2150)$&$1^{+}(1^{--})$&the reported values are in 
&&&\\
&& the range of $70-410$
&&&\\
\hline
\end{tabular}
\end{table}

\begin{figure}[H]
\centering
\includegraphics[width=0.7\textwidth]{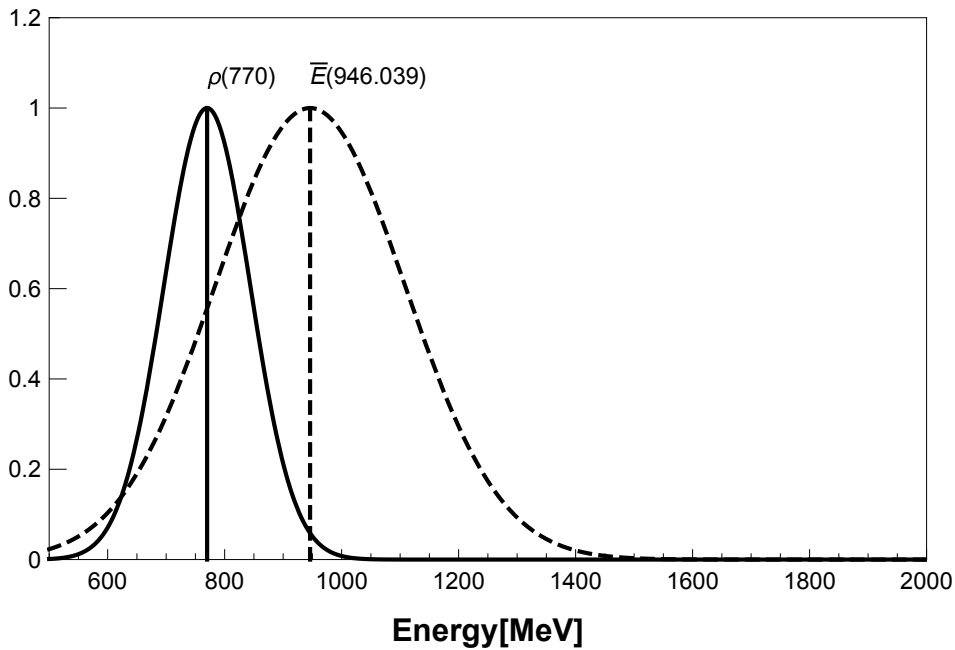}
\caption{Gaussian plot for $\rho$ and $\omega$ states. The experimental distribution (solid line) is compared to the calculated one (dashed lines)}
\label{GPrho1}
\end{figure}

Figure \ref{GPrho1} shows the gaussian plots for the energy and width of the experimental 
$(\rho(770),\omega(782))$ states and the theoretical multiplet, for which it is obtained  a mean energy of $\bar{E}(946.04)$. 
The calculated overlaps with the width of the experimental $(\rho(770),\omega(782))$ states and their centroid at $\bar{E}(946.04)$.

We continue the analysis of the widths by focussing on the range of energy of the reported  $(\rho(1450),\omega(1420))$ states, but since the experimental values of their widths are of the order of  $400MeV$ and $250MeV$ they overlap with the experimental $\rho(1570)$ and $\rho(1700)$ states. Thus, the  analysis in this case is extended over predicted states falling completely within the experimental bounds.This feature is well reproduced by the theoretical results, as it is shown in Table \ref{tab-m-1-}

\subsection{Mesons $1^+$}

\begin{figure}[H]
\centering
\includegraphics[width=0.7\textwidth]{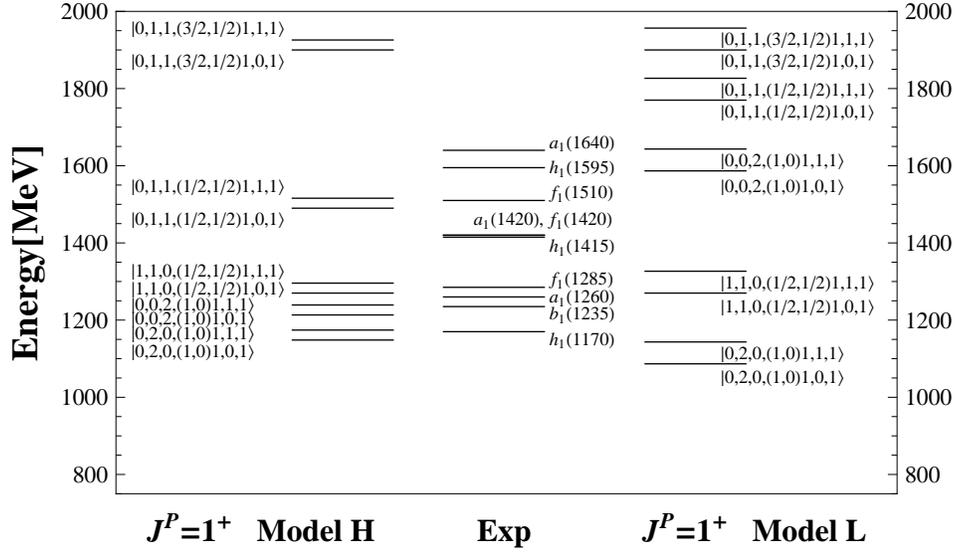}
\caption{Theoretical and experimental spectrum for $1^+$ meson states. }
\label{HLMesones1P}
\end{figure}

According to our classification  of the experimental states, we can
associate to the subspace of $1^+$ meson-like states
 six multiplets within the range of energies
$1100-1700$MeV (See Figure (\ref{HLMesones1P}). 
The first excited multiplets, for both negative and positive charge
conjugation, are located at higher energy than those obtained with the
model. However, the order is reproduced, since the first negative
charge conjugate state appears below the first excited positive charge conjugate state.

This subspace of the spectrum contains
states with positive and negative charge conjugation. The reference states are listed in Table 
\ref{tab-m-1++} and Table \ref{tab-m-1+-}, respectively.
\beqa
C & = & (-1)^{L+S}
~~~.
\eeqa

Looking at the charge conjugation properties, the spectrum presents some
order. Another characteristic of this subspace is that some widths are
very large, making the identification of the states rather difficult.
In order to make a reasonable comparison with the  model, we have to
accommodate the experimental states in the following form; $(h_1,b_1)$
as a multiplet and $(a_1,f_1)$ as another multiplet. This is because
the model can distinguish the angular momentum,  but not the isospin. Thus, we accommodate the
$(h_1(1170),b_1(1235))$ in a multiplet with negative charge
conjugation, and the $(a_1(1260),f_1(1285))$ in a multiplet with positive charge
conjugation. Then, the first excited multiplet with negative charge
conjugation will be the $(h_1(1415),?)$ one, and the first excited multiplet
with positive charge conjugation will be
$(a_1(1420),f_1(1420))$ pair.
The second excited multiplet with negative charge conjugation is
$(h_1(1595),?)$ and the second excited multiplet with positive charge
conjugation is the $(a_1(1640),f_1(1510))$ pair.
We have taken as multiplets $(h_1(1415),?)$ and
$(h_1(1595),?)$, although
the counterpart $b_1$ of isospin $I=1$ is missing. This is a recurrent feature at higher energies.

\begin{table}[H]
\centering
\caption{Experimental and theoretical values of the energy and width for mesons 
with $J^{PC}$ = $1^{+-}$ and isospin $I$.}
\label{tab-m-1+-}
\begin{tabular}{|c|c|c|c|c|c|}
\hline
E [MeV]&$I^{G}(J^{PC})$&width ([MeV])
&$| N_{0},N_{\frac{1}{2}},N_{1},(k_{1},k_{2}),L,S,J\rangle$
& $\bar{E}$[MeV]& $\Gamma $[MeV]\\ \hline

$h_{1}(1170)$&$0^{-}(1^{+-})$&$375$
&$| 0, 2, 0, ( 1, 0 ), 1, 0, 1\rangle$ & 1125.93 & 252.96 \\
$b_{1}(1235)$&$1^{+}(1^{+-})$&$142$
&$| 1, 1, 0, (\frac{1}{2}, \frac{1}{2}), 1, 0, 1\rangle$ & 1268.65 & 222.38 \\

$h_{1}(1415)$&$0^{-}(1^{+-})$&$90$
&&&\\

$h_{1}(1595)$&$0^{-}(1^{+-})$&$384$
&$| 0, 0, 2, ( 1, 0 ), 1, 0, 1 \rangle$& 1589.91 & 278.94 \\
&&&$| 0, 1, 1, (\frac{1}{2}, \frac{1}{2}), 1, 0, 1 \rangle$& 1762.37 &208.53 \\
\hline
\end{tabular}
\end{table}

\begin{table}[H]
\centering
\caption{Experimental energies of mesons
with $J^{PC}$ = $1^{++}$ and isospin $I$.}
\label{tab-m-1++}
\begin{tabular}{|c|c|c|c|c|c|}
\hline
E [MeV]&$I^{G}(J^{PC})$&width ([MeV])
&$| N_{0},N_{\frac{1}{2}},N_{1},(k_{1},k_{2}),L,S,J\rangle$
& $\bar{E}$[MeV]& $\Gamma$ [MeV]   \\ \hline

$a_{1}(1260)$&$1{-}(1^{++})$&estimated: $250-600$; its average: $420$
&$| 0, 2, 0, ( 1, 0 ), 1, 1, 1\rangle$ & 1156.55 & 147.11\\
$f_{1}(1285)$&$0^{+}(1^{++})$&$22.7$
&&&\\

$f_{1}(1420)$&$0^{+}(1^{++})$&$54.5$
&&&\\
$a_{1}(1420)$&$1^{-}(1^{++})$&$161$
&$| 1, 1, 0, (\frac{1}{2}, \frac{1}{2}), 1, 1, 1\rangle$ & 1326.13 & 119.31 \\

$f_{1}(1510)$&$0^{+}(1^{++})$&$73$
&&&\\
$a_{1}(1640)$&$1^{-}(1^{++})$&$254$
&$| 0, 0, 2, ( 1, 0 ), 1, 1, 1 \rangle$& 1644.84 &152.63 \\ \hline
\end{tabular}
\end{table}

\subsection{Mesons $2^-$}

\begin{figure}[H]
\centering
\includegraphics[width=0.7\textwidth]{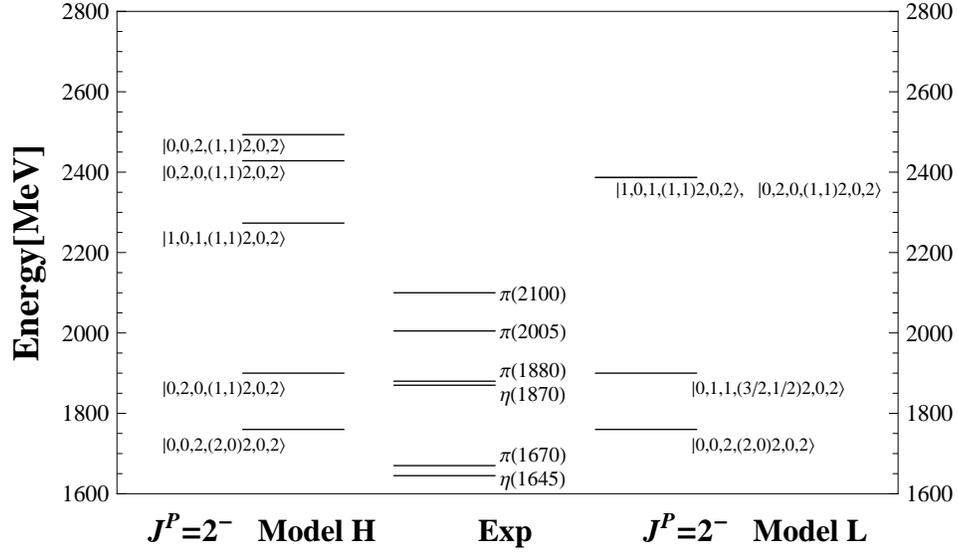}
\caption{Comparison between theoretical and experimental spectra for $J^P$ =
$2^{-}$ mesons.(see the captions to Figure  1 for further details)}
\label{HLMesones2M}
\end{figure}

The subspace of the meson spectrum with
quantum numbers $J^P=2^-$,
contains six observed states
within the range of energies $1645-2100$MeV.
They are $\pi$ and $\eta$ mesons. In order to make a
comparison with the results of the models we have grouped  the $\pi$'s
and $\eta$'s in multiplets. Therefore, we consider
$(\eta(1645),\pi(1670))$ as the lowest multiplet, then the
$(\eta(1870),\pi(1880))$ as the first excited multiplet, the
$(?,\pi(2005))$ as the second excited multiplet with the absence of
a $\eta$ -state and the $(?,\pi(2100))$  as the third excited
multiplet, again with the absence of the $\eta$ partner.
It is worth mentioning that the second and third multiplets have very
large widths (see Table \ref{tab-m-2-}),
making their energy identification
in some way dubious.
The other multiplets have considerable widths which could be the
reason why the $\pi$-states appear a little bit higher than their counterpart of the  $\eta$-states. As done with the other meson-like subspaces in Figure \ref{HLMesones2M} we show the correspondence between
 reference states and the data for each model spaces.

\begin{table}[H]
\centering \caption{Experimental energies for mesons with $J^P$ =
$2^{-}$ and isospin $I$.}
\label{tab-m-2-}
\begin{tabular}{|c|c|c|c|c|c|}
\hline
E [MeV]&$I^{G}(J^{PC})$& Width [MeV]
&$|N_{0},N_{\frac{1}{2}},N_{1},(k_{1},k_{2}),L,S,J\rangle$ & $\bar{E}$[MeV] & $\Gamma $[MeV] \\  \hline
$\eta_{2}(1645)$&$0^{+}(2^{-+})$&$181$
&&&\\
$\pi_{2}(1670)$&$1^{-}(2^{-+})$&$258$
&$|0, 0, 2, (2, 0), 2, 0, 2 \rangle$ & 1804.34 & 237.47  \\
&&&&&\\

$\eta_{2}(1870)$&$0^{+}(2^{-+})$&$225$
&$|0, 1, 1, (\frac{3}{2}, \frac{1}{2}), 2, 0, 2\rangle$ &  1893.12 & 281.25 \\
$\pi_{2}(1880)$&$1^{-}(2^{-+})$&$137$
&&&\\
&&&&&\\

$\pi_{2}(2005)$&$1^{-}(2^{-+})$&$370$
&&&\\
$\pi_{2}(2100)$&$1^{-}(2^{-+})$&$625$
&&&\\ \hline
\end{tabular}
\end{table}

\begin{figure}[H]\label{mesons2M}
\centering
\includegraphics[width=0.7\textwidth]{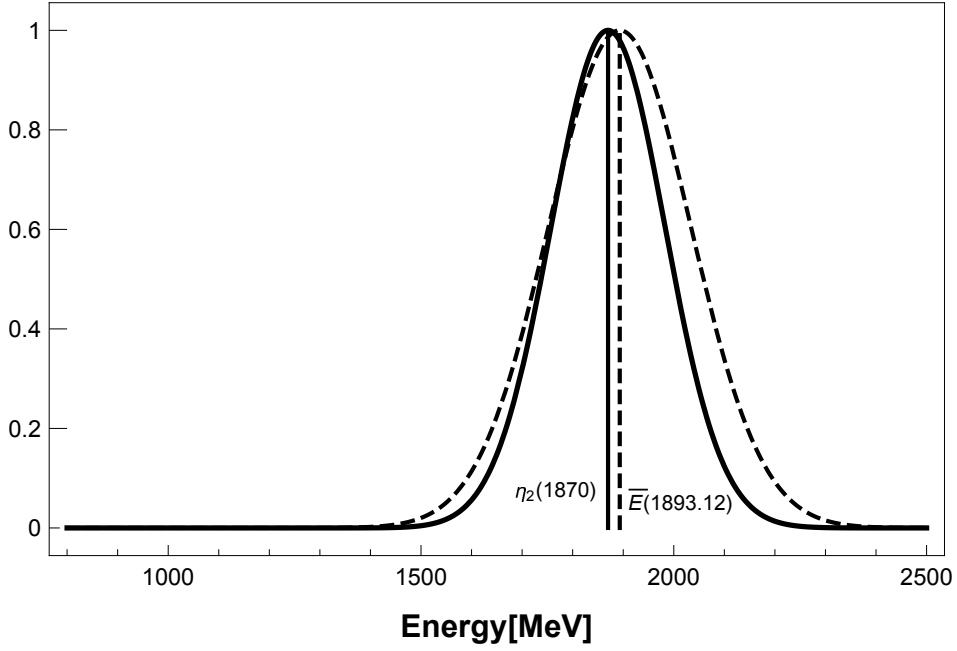}
\caption{Gaussian distribution for $J^P$ =
$2^{-}$ states. The experimental energy distribution (solid line) is compared to the theoretical values (dashed lines)} 
\end{figure}

As we have said we could expect four multiplets in this subspace
of the meson spectrum. 
The {\it Model-L} contains
exactly four multiplets. For these
multiplets we can see that at least the quark or the antiquarks should
occupy an SO(4) excited representation in order to have angular
momentum $L=2$ and that
for all of them the spin of the quark-antiquarks pair is
zero. These four multiplets, described by
the model, are located within a
range of energies between $1750-2400$ MeV. The two highest multiplets of the model, which
are degenerate,  are also in good agreement with the experimental
observations, since they fit within the range of the
large  experimental widths of the
multiplets $(?,\pi(2005))$  and $(?,\pi(2100))$, which are of the order of 370-625 MeV , making the experimentally observed states compatible with the reference states at approximately 2300 MeV.

\subsection{Mesons $2^+$}

\begin{figure}[H]
\centering
\includegraphics[width=0.7\textwidth]{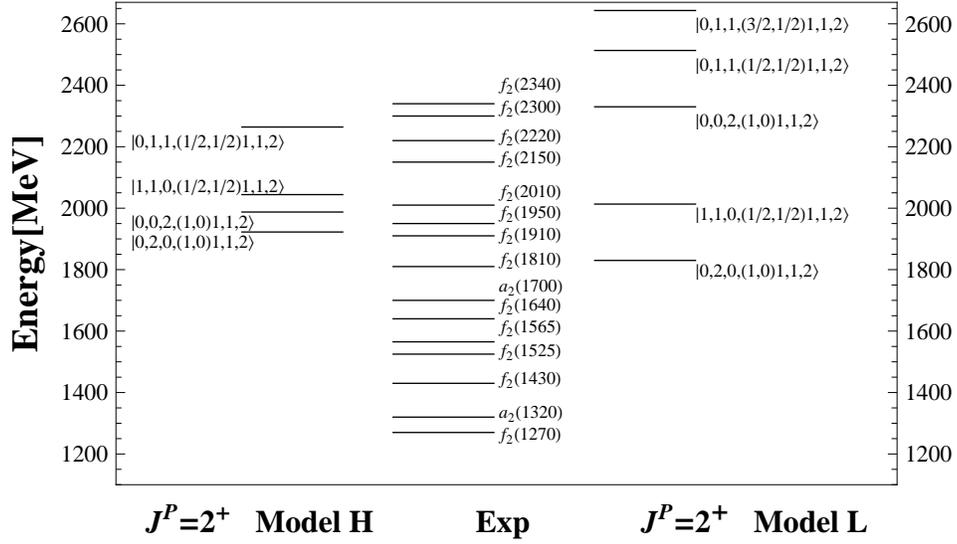}
\caption{Comparison between theoretical and experimental energies for $J^P=2^+$ mesons. The results are arranged as 
explained in the captions to Figure 1}
\label{HLMesones2P}
\end{figure}

The spectrum of mesons with quantum numbers $J^P=2^+$
is one of the most dense (see Figure \ref{HLMesones2P}). In the range of energies $1270-2340$MeV the experiments report
{\it thirteen} multiplets $(f_2,a_2)$, though some states
$a_2$ are missing.

This density of
states corresponds to a multiplet at every interval of nearly $100$MeV. This is a feature that the models cannot reproduce.

\begin{table}[H]
\centering
\caption{Experimental values for mesons
with $J^P$ = $2^{+}$ and isospin $I$.}
\label{tab-m-2+}
\begin{tabular}{|c|c|c|c|c|c|}
\hline
E [MeV]&$I^{G}(J^{PC})$&width ([MeV])
&$|N_{0},N_{\frac{1}{2}},N_{1},(k_{1},k_{2}),L,S,J\rangle$ & $\bar{E}$ [MeV] & $\Gamma$ [MeV] \\ \hline
$f_{2}(1270)$&$0^{+}(2^{++})$& average: $185.9$
&&&\\
$a_{2}(1320)$&$1^{-}(2^{++})$&$105$
&&&\\
$f_{2}(1430)$&$0^{+}(2^{++})$& $13-150$
&&&\\
$f_{2}'(1525)$&$0^{+}(2^{++})$&$86$
&&&\\
$f_{2}(1565)$&$0^{+}(2^{++})$&$122$
&&&\\
$f_{2}(1640)$&$0^{+}(2^{++})$&$99$
&&&\\
$a_{2}(1700)$&$1^{-}(2^{++})$&$258$
&&&\\
$f_{2}(1810)$&$0^{+}(2^{++})$&$197$
&&&\\

$f_{2}(1910)$&$0^{+}(2^{++})$&$167$
&$|0, 2, 0, (1, 0), 1, 1, 2\rangle$ &  1851.45 & 187.23 \\

$f_{2}(1950)$&$0^{+}(2^{++})$&$464$
&&&\\
$f_{2}(2010)$&$0^{+}(2^{++})$&$202$
&$|1, 1, 0, (\frac{1}{2}, \frac{1}{2}), 1, 1, 2 \rangle$ &  2012.49 & 156.03 \\

$f_{2}(2150)$&$0^{+}(2^{++})$&$152$
&&&\\
this observed state may && 
&&&\\
also be a $4^+$, $f_{J}(2220)$&$0^{+}(2^{++})$&$23$
&&&\\
$f_{2}(2300)$&$0^{+}(2^{++})$&$149$
&&&\\
$f_{2}(2340)$&$0^{+}(2^{++})$&$322$
&$|0, 0, 2, (1, 0), 1, 1, 2\rangle$ &  2332.27 & 198.40 \\
&&&$|0, 1, 1, (\frac{1}{2}, \frac{1}{2}), 1, 1, 2 \rangle$ &  2507.41 & 150.86 \\
&&&$|0, 1, 1, (\frac{3}{2},\frac{1}{2}), 1, 1, 2\rangle$ & 2627.41 & 145.94  \\ \hline
\end{tabular}
\end{table}

The available experimental information on the widths of the states belonging to
this meson subspace, does not seem to be very helpful,  since in this
very dense spectrum there are considerable large widths, see Table
\ref{tab-m-2+}.
The data can be arranged in multiplets
$(f_2(1270),a_2(1320))$,
$(f_2(1430),?)$,
$(f_2(1525),?)$,
$(f_2(1565),?)$,
$(f_2(1640),a_2(1700))$,
$(f_2(1810),?)$,
$(f_2(1910),?)$,
$(f_2(1950),?)$,
$(f_2(2010),?)$,
$(f_2(2150),?)$,
$(f_2(2220),?)$,
$(f_2(2300),?)$,
$(f_2(2340),?)$, and in good approximation they are equidistant multiplets.
It is worth to mention, that for this subspace we have not selected
any representative energy in order to fix the parameters of the Hamiltonian.
However, both ansatz gave energies in the range
of the experimental values, including their widths.

For  {\it Model-H} we expect to get a denser
spectrum, but in this subspace even the
{\it Model-H} does not
show that kind of density.
On the other hand,
the {\it Model-L} shows
a wider range of energies and
also one more multiplet than the {\it Model-H}.

\subsection{Final remarks about meson states and their widths.}

\begin{figure}[H]
\centering
\includegraphics[width=1.\textwidth]{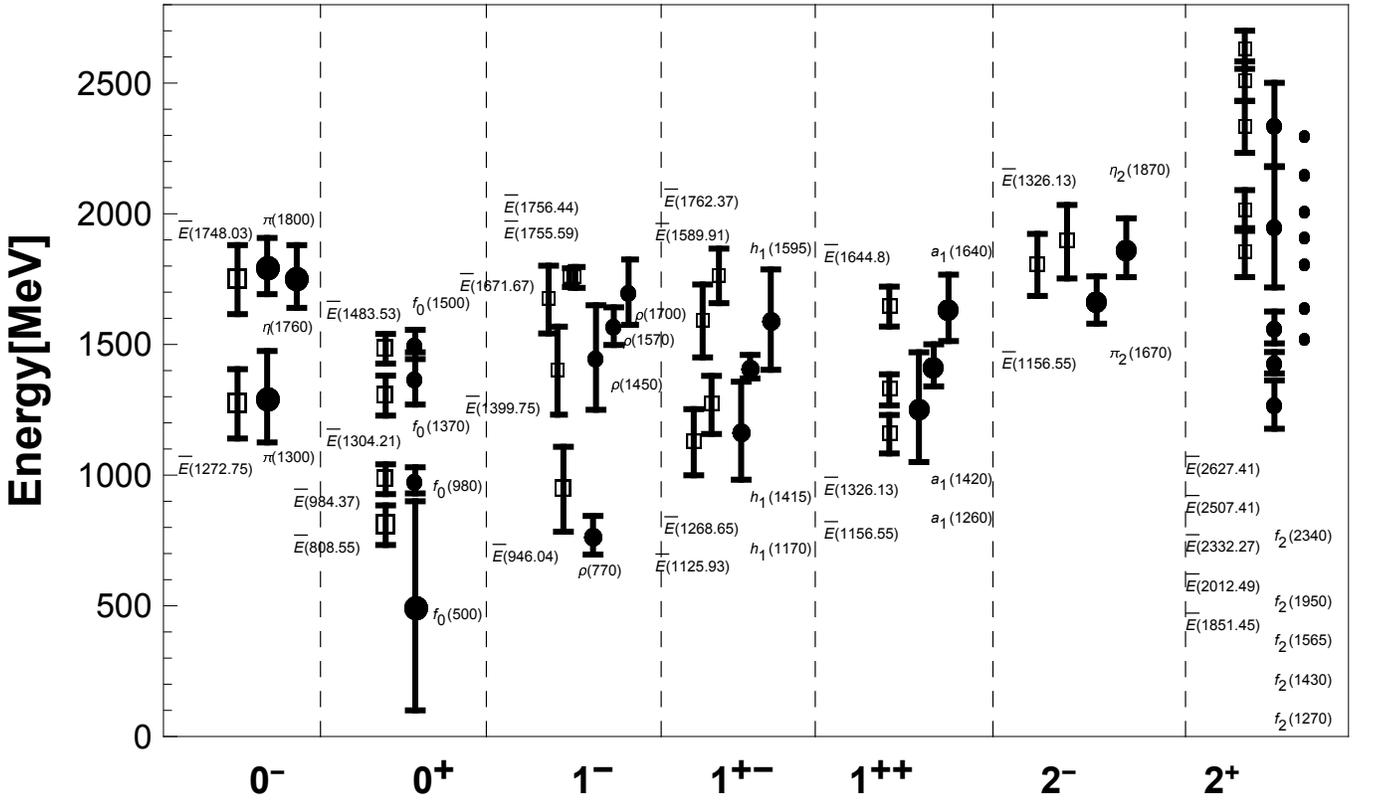}
\caption{Cumulative graph for the mean energy and width for meson states below 2.5 GeV. The experimental values (full circles) are compared to the theoretical values (open squares) obtained with Model L. The vertical bars are the experimetal and theoretical widths of the states. The dots which appear in set of $2^+$ states represent the experimentally found states, whose widths are very small when compared with their energies, as shown in Table \ref{tab-m-2+}.
}
\label{Widths}
\end{figure}

We have compared the experimental spectrum to the theoretical
one, looking at the energy scale and density of states, for mesons of various spins and parities. 
The theoretical classification of states within the SO(4) scheme, and the adopted parametrization of the Hamiltonian  gives a qualitatively good agreement with data, particularly for the linear model.
When the high energy sector of the meson spectrum is analyzed
(e.g., the $2^\pm$ states),
the theoretical spectrum agrees surprisingly well
in structure with the
experimental one, save that for $2^-$ states, where the
measured concentration of states in
a small range of energy is difficult to reproduce, though
the theory shows also the doublet structure. About the quality of the predictions for other spins and parities, 
the $0^\pm$ spectrum is fine but the $1^\pm$
spectrum shows some deficiencies, which in the case of the $1^+$ spectrum are due to the large density of states in a small energy range.
The difference between the energy of the first experimental and theoretical 
state
is also observed in the $2^\pm$ spectrum, but, in general, the theoretical values agree with the experimental ones after accounting for the observed widths of the states. Figure \ref{Widths} shows the compilation of theorerical and experimental results.

Finally, it is noted that the overall agreement to experiment is quite good, not only for the position of the states but also for their width. This is a surprise, considering the simplicity of the model, demonstrating that a well chosen algebraic model reflects the structure of the meson spectrum over a wide range of energies.

\section{Conclusions}
\label{conclusions}

In this work we have explored the validity of an effective model Hamiltonian which is applied to the description of meson states as superposition of quark-antiquark pairs belonging to the  SO(4) group representation. By working with the SO(4) basis we have 
associated to each set of meson states some reference states and mixed them up with background states in order to extract their widths. The systematic analysis of the results shows that the main features of the data are well reproduced, from a qualitative point of view and for some cases also quantitatively well reproduced, in spite of the rather schematic nature of the models used to calculate the spectra. We think that these results are indicative of a more fundamental assumption
about the role of symmetries in many body systems, as it was the case, for example, of correlations among nucleons in nuclei, where the group theoretical method has provided good explanations for observed properties like angular momentum and parity sequences. Work in is progress to extend the present formulation to include higher order correlations between quark-antiquark pairs.          
\section*{Acknowledgments}

P.O.H. acknowledges financial support from PAPIIT-DGAPA
(IN100421). O.C acknowledges the support of the CONICET 
and of the ANPCyT of Argentina.


\begin{thebibliography}{99}


\bibitem{Weinberg} S. Weinberg, {\it The Quantum Theory of Fields}
  (Vol. II, Cambridge University Press, 1996).


\bibitem{Lee-book} T. D. Lee, {\it Particle Physics and Introduction
to Field Theory} (Harwood Academic Publishers, New York, 1981).











\bibitem{Dudek1} 
J. J. Dudek, R. G. Edwards, M. J. Peardon, D. G. Richards and C. E. Thomas,
Phys. Rev D {\bf 82} (2010), 034508.


\bibitem{Dudek2}
  J. J. Dudek,
  Phys. Rev. D {\bf 84} (2011), 074023.

\bibitem{Dudek3} 
R. G. Edwards, J. J. Dudek, D. G. Richards and S. J. Wallace,
Phys. Rev. D {\bf 84} (2011), 074508.

\bibitem{Dudek4}
  J. J. Dudek and R. G. Edwards,
  Phys. Rev. D {\bf 85} (2012), 054016.

\bibitem{Dudek5}
  J. J. Dudek, R. G. Edwards, P. Guo and C. E. Thomas,
  Phys. Rev. D {\bf 88} (2013), 094505.

\bibitem{Bietenholz}
W. Bietenholz, Int. J. Mod. Phys. E 25 (2016) 1642008.


\bibitem{Bri18}
  R. A. Brice\~no, J. J. Dudek and R. D. Young,
  Rev. Mod. Phys. {\bf 90} (2018), 025001.

\bibitem{Edw20}
  R. G. Edwards, Proc. Sci. PoS(LATTICE2019)253 (2020).



\bibitem{Bashir}
A. Bashir et al., Comm. Theor. Phys. 58 (2012) 79.

\bibitem{DSE-baryons} 
G. Eichmann, H. Sanchis-Alepuz, R. Williams, R. Alkofer and C. S. Fischer, 
Prog. Part. Nucl. Phys. {\bf 91} (2016), 1.

\bibitem{DSE-Craig1}
  P.-L. Yin, C. Chen, G. Krein, C. D. Roberts, J. Segovia and S.-S. Xu,
  Phys. Rev. D {\bf 100} (2019), 034008.

\bibitem{DSE-Craig2}
  C. Chen, G. Krein, C. D. Roberts, S. M. Schmidt and J. Segovia,
  Phys. Rev. D {\bf 100} (2019), 054009.


\bibitem{Adam1996} 
A. Szczepaniak, E. S. Swanson, C. R. Ji and S. R.  Cotanch, 
Phys. Rev. Lett. {\bf 76}, 2011 (1996).

\bibitem{Llanes2000}   
F. J. Llanes-Estrada and S. R. Cotanch,  Phys. Rev. Lett. {\bf 84}, 1102 (2000).

\bibitem{Adam2001} A. P. Szczepaniak and E. Swanson, 
Phys. Rev. D {\bf  65},  025012 (2001).


\bibitem{Llanes2002} 
F. J. Llanes-Estrada and S. R. Cotanch,  Nucl. Phys. A   {\bf 697}, 303  (2002).


\bibitem{Hugo2004} C. Feuchter and H. Reinhardt, 
Phys. Rev. D {\bf    70}, 105021  (2004).

\bibitem{Hugo2011} H. Reinhardt, D. R. Campagnari and  A. P. Szczepaniak,  
Phys. Rev. D {\bf 84}, 045006 (2011).


\bibitem{Arturo2017}  D. A. Amor-Quiroz, T. Y\'epez-Mart\ii nez,  P. O. Hess, O. Civitarese,  and A. Weber,  Int. J. Mod. Phys. E  {\bf 26}, 1750082 (2017)




\bibitem{Ring} P. Ring and P. Schuck, {\it The Nuclear Many Body   Problem} (Springer, Heidelberg, 1980).




\bibitem{Bohr}
A. Bohr and B. R. Mottelson, Nuclear Structure V.I (World Scientific, Singapore, 1999).




\bibitem{sergio1}  
S Lerma H, S Jesgarz, P O Hess, O Civitarese, M Reboiro. Phys. Rev. C 66, 045207 (2002).

\bibitem{sergio2}

 S Lerma H, S Jesgarz, P O Hess, O Civitarese, M Reboiro. Phys. Rev. C 67, 055209
(2003).

\bibitem{sergio3}
 S Lerma H, S Jesgarz, P O Hess, O Civitarese, M Reboiro. Phys. Rev. C 67, 055210 (2003).


\bibitem{sergio4}
S Lerma H, S Jesgarz, P O Hess, O Civitarese, M Reboiro. Phys. Rev. C 70, 035208 (2004).




\bibitem{Hess2006} P. O. Hess and A. P. Szczepaniak, Phys. Rev. C {\bf 73},  025201 (2006).

\bibitem{Yepez2010}  T. Y\'epez-Mart\ii nez, P. O. Hess,  A. P. Szczepaniak and O. Civitarese, Phys. Rev. C {\bf 81}, 045204 (2010).

\bibitem{SO4-1} T. Yepez-Martinez, O. Civitarese and P. O. Hess, 
Int. J. Mod. Phys. E {\bf 25}, 1650067  (2016).

\bibitem{SO4-2} T. Yepez-Martinez, O. Civitarese and P. O. Hess, 
Int. J. Mod. Phys. E {\bf 26}, 1750012 (2017).

\bibitem{SO4-3}
T. Y\'epez-Mart\ii nez, O. Civitarese and P. O. Hess, 
Int. J. Mod. Phys. E {\bf 27},  1850001 (2018)







\bibitem{SO4-4} 
U. I. Ramirez-Soto, O. A. Rico-Trejo, T. Y\'epez-Mart\'inez,
P. O. Hess, A. Weber and O. Civitarese, 
J. Phys. G: Nucl. Part. Phys. {\bf 48}, 085013 (2021).





\bibitem{Rujula} 
De R\'ujula A, Georgi H and Glashow S L 1975 Phys. Rev. D 12 147



\bibitem{booklet2020} P. A. Zyla et al. Extracted from the Particle Data Group, Prog.Theor.Exp. Phys. 2020, 083C01(2020)





\bibitem{Greiner}
Greiner W and Muller B 1994 Quantum Mechanics: Symmetries (Heidelberg: Springer)


\end{thebibliography}
\end{document}